\documentclass[aps,prc,showpacs,twocolumn,superscriptaddress,floatfix]{revtex4}

\usepackage[dvips]{graphicx}

\begin{document}

\title{Particle production in central Pb+Pb collisions at 158 $A$ GeV/$c$}

\author{I.G.~Bearden}
  \affiliation{Niels Bohr Institute, DK-2100, Copenhagen, Denmark}
\author{H.~B{\o}ggild}
  \affiliation{Niels Bohr Institute, DK-2100, Copenhagen, Denmark}
\author{J.~Boissevain}
  \affiliation{Los Alamos National Laboratory, Los Alamos, New Mexico 87545}
\author{P.H.L.~Christiansen}
  \affiliation{Niels Bohr Institute, DK-2100, Copenhagen, Denmark}
\author{L.~Conin}
  \affiliation{Nuclear Physics Laboratory of Nantes, 44072 Nantes, France}
\author{J.~Dodd}
  \affiliation{Columbia University, New York, New York 10027}
\author{B.~Erazmus}
  \affiliation{Nuclear Physics Laboratory of Nantes, 44072 Nantes, France}
\author{S.~Esumi}
  \altaffiliation{Now at Univ. of Tsukuba, Tsukuba, Ibaraki 305-8571, Japan}
  \affiliation{Hiroshima University, Higashihiroshima, Hiroshima 739-8526, Japan}
\author{C.W.~Fabjan}
  \affiliation{CERN, CH-1211 Geneva 23, Switzerland}
\author{D.~Ferenc}
  \altaffiliation{Now at Univ. of California, Davis, CA 95616}
  \affiliation{Rudjer Bo{\u{s}}kovi{\'c} Institute, Zagreb, Croatia}
\author{D.E.~Fields}
  \altaffiliation{Now at Univ. of New Mexico, Albuquerque, NM 87185}
  \affiliation{Los Alamos National Laboratory, Los Alamos, New Mexico 87545}
\author{A.~Franz} 
  \altaffiliation{Now at BNL, Upton, NY 11973}
  \affiliation{CERN, CH-1211 Geneva 23, Switzerland}
\author{J.J.~Gaardh{\o}je}
  \affiliation{Niels Bohr Institute, DK-2100, Copenhagen, Denmark}
\author{M.~Hamelin}
  \affiliation{Cyclotron Institute, Texas A\&M University, College Station, Texas 77843}
\author{A.G.~Hansen}
  \altaffiliation{Now at LANL, Los Alamos, NM 87545}
  \affiliation{Niels Bohr Institute, DK-2100, Copenhagen, Denmark}
\author{O.~Hansen}
  \affiliation{Niels Bohr Institute, DK-2100, Copenhagen, Denmark}
\author{D.~Hardtke}
  \altaffiliation{Now at LBNL, Berkeley, CA 94720}
  \affiliation{Ohio State University, Columbus, Ohio 43210}
\author{H.~van~Hecke}
  \affiliation{Los Alamos National Laboratory, Los Alamos, New Mexico 87545}
\author{E.B.~Holzer}
  \affiliation{CERN, CH-1211 Geneva 23, Switzerland}
\author{T.J.~Humanic}
  \affiliation{Ohio State University, Columbus, Ohio 43210}
\author{P.~Hummel}
  \affiliation{CERN, CH-1211 Geneva 23, Switzerland}
\author{B.V.~Jacak}
  \affiliation{State University of New York, Stony Brook, New York 11794}
\author{K.~Kaimi}
  \altaffiliation{Deceased}
  \affiliation{Hiroshima University, Higashihiroshima, Hiroshima 739-8526, Japan}
\author{M.~Kaneta}
  \altaffiliation{Now at LBNL, Berkeley, CA 94720}
  \affiliation{Hiroshima University, Higashihiroshima, Hiroshima 739-8526, Japan}
\author{T.~Kohama}
  \affiliation{Hiroshima University, Higashihiroshima, Hiroshima 739-8526, Japan}
\author{M.~Kopytine}
  \altaffiliation{On unpaid leave from P.N.Lebedev Physical Institute, Russian Academy of Sciences}
  \altaffiliation{Now at Kent State Univ., Kent, OH 44242} 
  \affiliation{State University of New York, Stony Brook, New York 11794}
\author{M.~Leltchouk}
  \affiliation{Nuclear Physics Laboratory of Nantes, 44072 Nantes, France}
\author{A.~Ljubi{\u{c}}i{\'c},~Jr.}
  \altaffiliation{Now at BNL, Upton, NY 11973}
  \affiliation{Rudjer Bo{\u{s}}kovi{\'c} Institute, Zagreb, Croatia}
\author{B.~L{\"o}rstad}
  \affiliation{University of Lund, S-22362 Lund, Sweden}
\author{N.~Maeda}
  \affiliation{Hiroshima University, Higashihiroshima, Hiroshima 739-8526, Japan}
\author{R.~Malina}
  \affiliation{CERN, CH-1211 Geneva 23, Switzerland}
\author{L.~Martin}
  \affiliation{Nuclear Physics Laboratory of Nantes, 44072 Nantes, France}
\author{A.~Medvedev}
  \affiliation{Columbia University, New York, New York 10027}
\author{M.~Murray}
  \affiliation{Cyclotron Institute, Texas A\&M University, College Station, Texas 77843}
\author{H.~Ohnishi}
  \altaffiliation{Now at RIKEN, Wako, Saitama 351-0198, Japan}
  \affiliation{Hiroshima University, Higashihiroshima, Hiroshima 739-8526, Japan}
\author{G.~Paic}
  \affiliation{CERN, CH-1211 Geneva 23, Switzerland}
  \affiliation{Ohio State University, Columbus, Ohio 43210}
\author{S.U.~Pandey}
  \affiliation{CERN, CH-1211 Geneva 23, Switzerland}
\author{F.~Piuz}
  \affiliation{CERN, CH-1211 Geneva 23, Switzerland}
\author{J.~Pluta}
  \altaffiliation{Now at Institute of Physics, Warsaw Univ. of Technology, Koszykowa 75,00-662, Warsaw, Poland}
  \affiliation{Nuclear Physics Laboratory of Nantes, 44072 Nantes, France}
\author{V.~Polychronakos}
  \affiliation{Brookhaven National Laboratory, Upton, New York 11973}
\author{M.~Potekhin}
  \affiliation{Columbia University, New York, New York 10027}
\author{G.~Poulard}
  \affiliation{CERN, CH-1211 Geneva 23, Switzerland}
\author{D.~Reichhold}
  \altaffiliation{Now at Creighton Univ., Omaha, NE 68178}
  \affiliation{Ohio State University, Columbus, Ohio 43210}
\author{A.~Sakaguchi}
  \altaffiliation{Now at Osaka Univ., Toyonaka, Osaka 560-0043, Japan}
  \affiliation{Hiroshima University, Higashihiroshima, Hiroshima 739-8526, Japan}
\author{J.~Schmidt-S{\o}rensen}
  \affiliation{University of Lund, S-22362 Lund, Sweden}
\author{J.~Simon-Gillo}
  \altaffiliation{Now at U.S. Department of Energy, Germantown, MD, 20874-1290}
  \affiliation{Los Alamos National Laboratory, Los Alamos, New Mexico 87545}
\author{W.~Sondheim}
  \affiliation{Los Alamos National Laboratory, Los Alamos, New Mexico 87545}
\author{M.~Spegel}
  \affiliation{CERN, CH-1211 Geneva 23, Switzerland}
\author{T.~Sugitate}
  \affiliation{Hiroshima University, Higashihiroshima, Hiroshima 739-8526, Japan}
\author{J.P.~Sullivan}
  \affiliation{Los Alamos National Laboratory, Los Alamos, New Mexico 87545}
\author{Y.~Sumi}
  \altaffiliation{Now at Hiroshima International Univ., Kamo, Hiroshima 724-0695, Japan}
  \affiliation{Hiroshima University, Higashihiroshima, Hiroshima 739-8526, Japan}
\author{W.J.~Willis}
  \affiliation{Columbia University, New York, New York 10027}
\author{K.L.~Wolf}
  \altaffiliation{Deceased}
  \affiliation{Cyclotron Institute, Texas A\&M University, College Station, Texas 77843}
\author{N.~Xu}
  \altaffiliation{Now at LBNL, Berkeley, CA 94720}
  \affiliation{Los Alamos National Laboratory, Los Alamos, New Mexico 87545}
\author{D.S.~Zachary}
  \affiliation{Ohio State University, Columbus, Ohio 43210}

\collaboration{NA44 collaboration}
\noaffiliation

\date{\today}

\begin{abstract}
  The NA44 experiment has measured single particle inclusive spectra for charged pions, kaons, and protons as a function of transverse mass near mid-rapidity in 158 A GeV/$c$ Pb+Pb collisions.
  From the particle mass dependence of the observed $m_T$ distributions, we are able to deduce a value of about 120 MeV for the temperature at thermal freeze-out.
  From the observed ratios of the rapidity densities, we find values of the chemical potentials for light and strange quarks and a chemical freeze-out temperature of approximately 140 MeV.
\end{abstract}


\maketitle

\section{INTRODUCTION}

  The goal of heavy-ion collision research at ultra-relativistic energies is to understand the fundamental properties of nuclear matter under extreme conditions of high particle and energy density.
  When these densities and temperatures in heavy-ion collisions are sufficiently high, quarks should no longer be confined in hadrons but instead move freely through the entire volume.
  The colliding system expands and cools, driven by the large internal pressure generated by the strong interaction of the constituents.
  The quarks eventually become confined again, forming hadrons that can be detected by experiments.
  The hadron yields and momentum distributions provide information about the temperature, density, and dynamics of the later stages of the collision.
  From this inferences about the conditions at earlier stages can be made.~\cite{QM1999_2001}

  The multiplicity of the produced hadrons ranges from several hundred to a few thousand in relativistic heavy-ion collisions at $\sqrt{s_{NN}}\approx$20 GeV, and their rapidity density cannot be explained by simple superposition of $p$+$p$ collisions, as final state rescattering may play an important role.
  The transverse momentum distributions of the hadrons display a Boltzmann-like shape, inspiring a thermodynamical interpretation of the momentum distributions.
  \begin{figure*}
    \includegraphics[width=17.5cm]{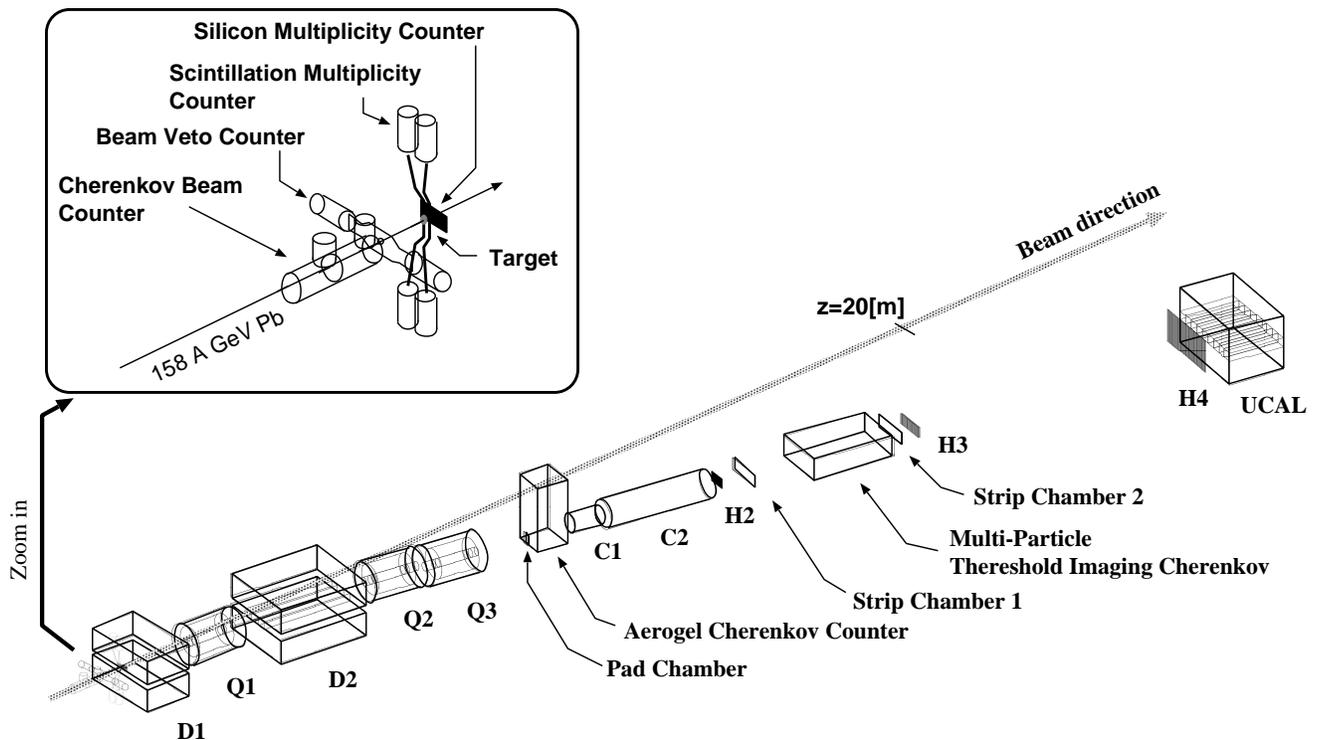}
    \caption{\label{fig:spect}
             The schematic view of NA44 spectrometer.}
  \end{figure*}

  In this report, we present invariant cross sections of $\pi^{\pm}$, $K^{\pm}$, $p$, and $\bar{p}$, as a function of transverse mass near mid-rapidity in central 158 A GeV/$c$ Pb+Pb collisions.
  NA44 has previously reported the particle yields for pions, kaons, and protons in $p$+A to S+A collisions~\cite{PR_C57_1998_837,PR_C59_1999_328}, kaon production in Pb+Pb~\cite{PL_B471_1999_6}, and antiproton and antideuteron production in Pb+Pb~\cite{prl_85_2000_2681}.
  Because of the excellent particle identification and systematic measurements from small to large systems, NA44 is well-suited to study the thermal and chemical properties of the hadronic system at freeze-out.
  The dependence of the thermal parameters on the colliding system size will also be discussed.

\section{NA44 EXPERIMENT}
  The NA44 focusing spectrometer is designed to measure one and two particle momentum distributions of charged hadrons near mid-rapidity~\cite{PL_B302_1993_510}.
  A schematic view of the spectrometer is shown in Fig.~\ref{fig:spect}.
  The target is a lead disk of 2 mm thickness (3.4\% interaction probability for a lead beam).
  A Cherenkov gas beam counter (CX) ~\cite{NIM_A346_1994_132} and a beam veto counter (CX-veto) are placed upstream of the target.
  CX selects single beam particles (rejecting events with multiple simultaneous beam particles) up to a rate of $2\times10^6$ ions/second while the CX-veto signals the presence of any beam halo.
  A scintillation multiplicity counter (T0) is 10 mm downstream of the target, and produces a pulse height proportional to the number of charged particles impinging upon it.
  T0 covers 20\% of the azimuthal angle and has a pseudo-rapidity range of 0.6${\leq}$${\eta}$${\leq}$3.3.
  This contains mid-rapidity ($\mbox{y}_{mid}=$2.9) for the Pb+Pb collision.
  T0 is used to select collision centrality at the trigger level.

  Two dipole magnets (D1 and D2) analyze the particle momenta.
  Three super conducting quadrupole magnets (Q1, Q2, and Q3) focus the particles.
  The PID/tracking detectors measure hit positions after the magnetic analysis, and consist of strip chambers and scintillation hodoscopes, along with two threshold gas Cherenkov counters, C1 and C2.
  Three hodoscopes, H2, H3, and H4 consist of 60, 50, and 60 scintillation counters, respectively.
  Only H2 and H3 are used in this analysis.
  The hodoscopes provide position information for tracking, with vertical hit information determined from photomultiplier tubes on each end of every counter.
  The hodoscopes also measure time-of-flight for hadron identification~\cite{NIM_A287_1990_287}.
  One pad chamber (PC) and two strip chambers (SC1 and SC2) provide 0.3 mm position resolution for track reconstruction.
  C1 and C2 contain Freon-12 at 2.7 atm and N$_2$ at 1.3 atm respectively, and identify particles at the trigger level.
  The trigger requires a valid beam, a high multiplicity detected in T0 (central collision), at least one track in each hodoscope, and the absence of an electron (all runs) or pion (for $K$ and $p$ runs).

  The timing signal from CX is used for the time-of-flight (TOF) start with a resolution of 35 ps.
  The average of the top and bottom PMT of the hodoscope is used for the stop signal.
  The overall TOF resolution is approximately 100 ps and 80 ps for H2 and H3, respectively.

  \begin{figure*}
    \includegraphics[width=15cm]{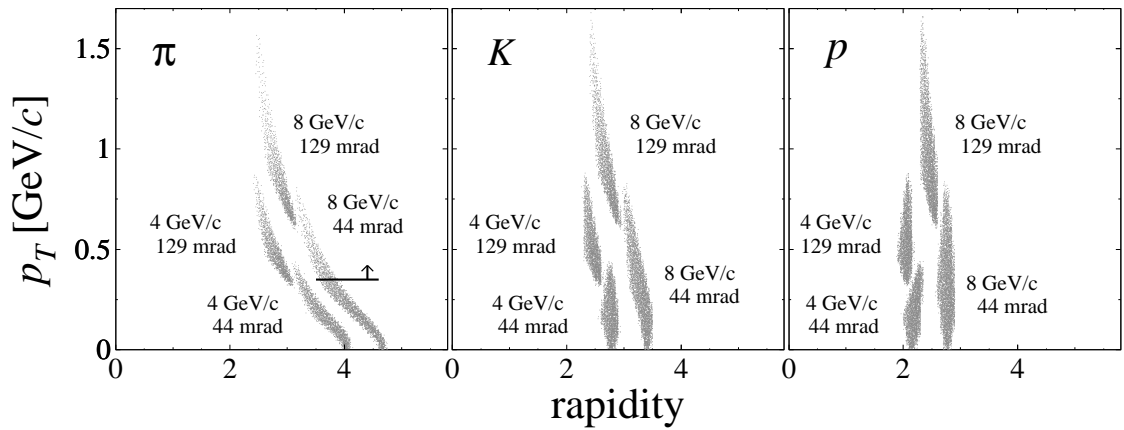}
    \caption{\label{fig:ypt}
             The $\pi$, $K$, and $p$ acceptance in rapidity and $p_T$.
             In this analysis, pions at the 8 GeV/$c$ 44 mrad angle setting are selected with $p_T$$>$0.35 GeV/$c$.
            }
  \end{figure*}
  \begin{figure*}
    \includegraphics[width=15cm]{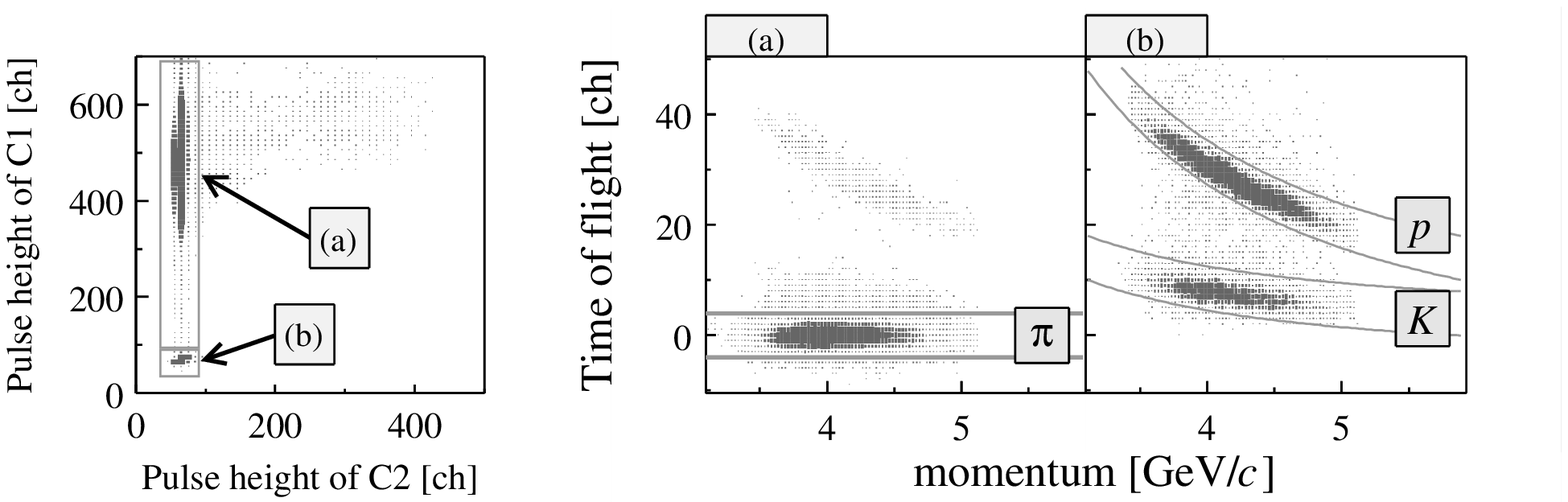}
    \caption{\label{fig:TOF2p}
             An example of particle identification in the 4 GeV/$c$ setting.
             The contour plot shows pulse heights of C1 and C2 (left).
             The region (a) is electron veto and pion required by C1 and C2 pulse heights.
             The region (b) is electron and pion veto by C1 and C2. 
             The right-hand plots are plots of the time-of-flight vs. track momentum for events inside the boxes labeled (a) and (b) in the left-hand plot.
             The pair of plots on the right are TOF at H3 as a function of momentum with C1 and C2 cuts.
             The curves show $\pm$2.5 sigma of TOF resolution from the peak.
             Particles between the curves are selected.
            }
  \end{figure*}

  Two magnetic field settings of the spectrometer are used, yielding nominal momentum of 4 GeV/$c$ ($3.2{\le}p{\le}5.2$ GeV/$c$) and 8 GeV/$c$ ($6.0{\le}p{\le}9.8$ GeV/$c$).
  Two angular settings, 44 mrad and 129 mrad are used to give a large range in transverse momentum ($p_T$).
  Combining these settings yields transverse momenta from $p_T=0$ to 0.9 GeV/$c$ (0 to 1.6 GeV/$c$) for 4 GeV/$c$ (8 GeV/$c$) setting.
  Figure~\ref{fig:ypt} shows the NA44 acceptance in $p_T$ and rapidity for pions, kaons, and protons.
  The NA44 acceptance includes mid-rapidity ($\mbox{y}_{mid}$=2.9) for Pb+Pb collisions at 158 GeV/$c$ per nucleon.

  The invariant cross section is determined by the following analysis procedures:
  track and momentum reconstruction, particle identification of tracks, a selection of high multiplicity events, an acceptance correction via a Monte-Carlo simulation, normalization by summing of beam scalers, and multiplicative correction factors for cuts.

  The hit positions on PC, H2, H3, SC1, and SC2 were used to reconstruct tracks by fitting them with a straight line using a $\chi^2$-minimization method.
  As is clear from Fig.~\ref{fig:spect}, the tracking detectors lie outside the magnetic field.
  The momentum resolution $\delta p/p$ is 0.2\% for all particle species.

  \begin{table}
    \caption{\label{tbl:sta}
             The number of particles selected by the cuts for each spectrometer setting.
             The event fraction is top $3.7\pm0.1\%$.}
    \begin{ruledtabular}
    \begin{tabular}{ccrrrrrr}
      nominal  & angle&      &      &      &      &      &      \\
      momentum &[mrad] & \multicolumn{1}{c}{$\pi^+$} 
                    & \multicolumn{1}{c}{$\pi^-$}
                             & \multicolumn{1}{c}{$K^+$}
                                    & \multicolumn{1}{c}{$K^-$}
                                             & \multicolumn{1}{c}{$p$}
                                            & \multicolumn{1}{c}{$\bar{p}$} \\
     \hline
     4 GeV/$c$ &  44 &  2713&  3839&  5150&  5814&  3644&  428 \\
               & 129 & 11498& 11163&  4238& 12222&  6577& 1663 \\
     8 GeV/$c$ &  44 &  1272&   992& 14689& 12563&  6215&  711 \\
               & 129 &  5915&  6785& 22614& 26578& 41682& 5834
    \end{tabular}
    \end{ruledtabular}
  \end{table}
  Particle identification uses a combination of pulse heights in C1 and C2 with time-of-flight from CX to the hodoscopes.
  The Cherenkov threshold momenta of C1(C2) for $e$, $\pi$, $K$, and $p$ are 0.014, 2.5, 8.9, and 16.9 (0.020, 5.3, 18.3, and 34.8) GeV/$c$, respectively.
  In the 4 GeV/$c$ spectrometer setting, pions are separated from $K$ and $p$ by C1 and C2, and then $K$/$p$ separation is done by TOF.
  In the 8 GeV/$c$ setting, $e$ and $\pi$ are vetoed by C2, and then $K$/$p$ separation is done by C1.
  The $e$/$\pi$ separation is not able to be done by C1 and C2 in the 8GeV/$c$ setting, however the contamination from electrons is negligible for $p_T$$>$0.35 GeV/$c$.
  The statistics for each particle are shown in Table~\ref{tbl:sta}.

  Because it has a longer flight path in addition to its better timing resolution, H3 rather than H2 provides the primary TOF information.
  Figure~\ref{fig:TOF2p} shows an example of particle identification in the 4 GeV/$c$ setting.
  The left-hand side shows a scatter plot of pulse heights in C1 and C2.
  Events containing an electron fire C2 -- these events are rejected.
  The events selected by the box labeled (a) and (b) on the left-hand side of figure are shown on plots of TOF as a function of momentum on the right-hand side of the figure.
  Particles between the curves are selected.
  The contamination for each particle species is less than 3\% in both momentum settings.
  \begin{figure}
    \includegraphics[width=8.5cm]{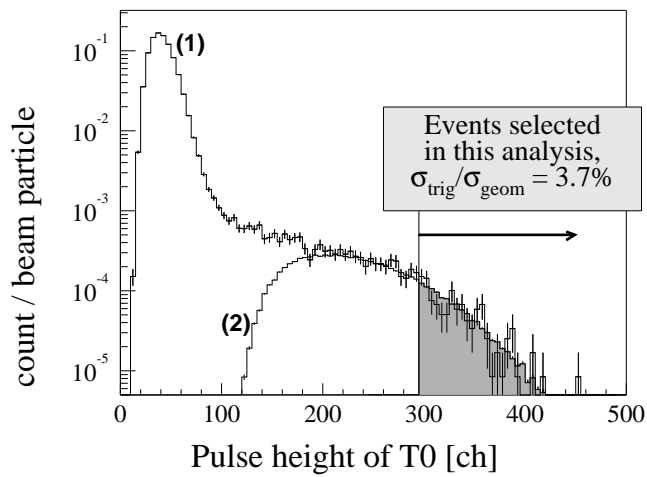}
    \caption{\label{fig:T0}
             The pulse height distribution of the T0 multiplicity counter for valid beam events (1) and events that satisfy the T0 centrality trigger (2).
             The shaded region is used for the analysis.
            }
  \end{figure}

  We trigger on central collisions using the interaction counter, T0, which measures the total amount of light produced by charged particles crossing its scintillators.  
  Figure~\ref{fig:T0} shows the pulse height distribution of the T0 for valid beam events (1) and central triggers (2).
  The shaded region is used for the analysis.
  The width of the valid  beam distribution comes from the production of delta electrons.
  Since T0 covers only 20\% of the azimuthal angle, it is not possible to count all low multiplicity Pb+Pb interactions and we cannot measure the Pb+Pb interaction cross section.
  Therefore we scale the measured $p$+Pb cross section \cite{ParticleData} by a geometry factor: this fraction is denoted by $\lambda$.
  The value of $\lambda$ is obtained by scaling the interaction probability for $p$+Pb collisions (the nuclear interaction length) to the interaction probability for Pb+Pb collisions by a geometry factor of $(208^{\frac{1}{3}}+208^{\frac{1}{3}})^2$/$(1^{\frac{1}{3}}+208^{\frac{1}{3}})^2$, i.e., $\lambda$=2.93.
  The nuclear interaction length of Pb is 194 g/cm$^2$ and density of Pb is 11.35 g/cm$^3$.
  Therefore, the interaction probability of $p$+Pb of 2 mm Pb target is $0.2/(194/11.35)=0.0117$.
  The total Pb interaction probability is 3.4\%$(=\lambda\times0.0117)$ for 2 mm Pb target.

  We then find that the shaded region in Fig.~\ref{fig:T0} corresponds to $3.7 \pm 0.1$\% (statistical only) of the Pb+Pb interaction cross section.
  This region corresponds to a region of impact parameter ${\langle}b{\rangle}$$\simeq<$3 (RMS$\simeq$1) fm, estimated by applying the T0 acceptance cut to the multiplicity distribution from RQMD (v2.3)~\cite{PR_C52_1995_3291}.
  We note that the T0 covers only 20\% of the azimuthal angle, therefore the event centrality selected by T0 has a weak sensitivity to multiplicity, that is, to impact parameter.

\section{ANALYSIS}

  In order to construct the invariant cross section, the effect of the acceptance is estimated by a Monte-Carlo (MC) simulation based on TURTLE~\cite{TURTLE}, which passes generated tracks through the detectors.
  The acceptance, reconstruction efficiencies, momentum resolution, and loss of pions and kaons due to in-flight-decays are estimated by comparing generated tracks with reconstructed tracks.
  The momentum spectrum and rapidity distribution for each particle species in Pb+Pb collisions is taken from RQMD (v2.3).
  The Monte-Carlo events were analyzed with the tracking programs exactly as applied to the physics events.
  The corrections are given by the ratio of the generated yield to the reconstructed yield as a function of $m_T$ inside the rapidity acceptance.
  The absolute cross section as a function of $m_T$ in a rapidity range ${\mit\Delta}y$ is given by simplify this equation.
    \begin{eqnarray*}
      \left.
        \frac{1}{\sigma} \frac{E d^3 \sigma}{dp^3}
      \right|_{{\mit\Delta}y}
      &
      =
      &
        \frac{1}{N_{\text{event}}} \frac{1}{2\pi}
        \frac{dN}{m_T dm_T {\mit\Delta}y} \\
      & = &
      \frac{1}{N_{\text{event}}} \frac{dn}{dm_T {\mit\Delta}y}
      \times
      \frac{
            \frac{1}{2\pi} \frac{1}{m_T} \frac{d\breve{n}_{mc}}{dm_T {\mit\Delta}y}
           }
           {
            \frac{d\breve{n}_r}{dm_T {\mit\Delta}y}
           }
    \end{eqnarray*}
  where $n$ is the number of measured tracks, $N_{\text{event}}$ is the number of collisions, $\breve{n}_{mc}$ is the number of particles as input of the MC simulation and $\breve{n}_r$ is the number of tracks reconstructed in the simulation.

  Secondary hadronic interactions of protons and antiprotons in the spectrometer material 
  does not distort the shape of the momentum distributions but reduces the yield\cite{PR_C57_1998_837}.
  The correction factors are estimated to be 1.11 for protons and 1.17 for antiprotons, respectively.
  
  The corrected momentum distribution is then normalized using the number of beam particles, the interaction probability, the fraction of interactions satisfying the trigger, the measured live time of the data-acquisition system, and losses due to particle identification cuts.
 
  Separation of protons and kaons from pions is done via Cherenkov veto trigger, but in the process some protons and kaons are rejected in events that also include a pion.
  The veto factor (i.e., the correction for this loss) for each particle and momentum setting is estimated by comparing the number of the particles with no Cherenkov veto to the number of particles with Cherenkov veto, using real data without the Cherenkov veto trigger.
  Since the veto factor depends on relative particle ratios in spectrometer, it must be independently estimated for each momentum and angle setting.
  
  The transverse mass distribution is obtained from the two angle settings (44 mrad and 129 mrad).
  The data at 129 mrad have smaller (typically one half to one third) systematic errors in the absolute normalization than those at 44 mrad.
  Therefore, in this analysis, the data at 44 mrad were scaled to match those at 129 mrad in the $m_T$ region where they overlap.
  The distributions from both settings were fit by exponentials with the same inverse slope parameter.
  The dN/dy is the sum of the normalized cross section in each measured $m_T$ bin within the acceptance plus an extrapolation that uses the fitted exponential distribution beyond the measurement region.
  This extrapolation contributes less than 1\% of the yield for pions  but up to 7\% for protons with the 4 GeV/$c$ setting.

\section{SYSTEMATIC ERRORS}

  The inverse slope parameters are extracted by fitting the experimental distribution with a single $m_T$ exponential function (see Eq.~(\ref{eqn:mt_exponential}) in the section~\ref{section:results}).
  Systematic errors on the inverse slope parameters are estimated by dividing the data into two $p_T$ ranges for the fits.
  The systematic errors are shown with the inverse slope parameters in Table~\ref{tbl:invslope_dndy}.

  The absolute normalization procedure results in significant uncertainties in dN/dy.
  Contributions include uncertainties on the following; the data-acquisition live time ($5\%$), the stability of the T0 pulse height used to select the event fraction ($2\%$), the Cherenkov veto factor ($<$$5\%$), the correction for particle loss due to quality cuts ($<$$5\%$), and the extrapolation beyond the measurement region ($<$$1{\sim}7\%$).
  The last three vary with particle species and spectrometer setting.
  We assume that the errors are uncorrelated and so add them in quadrature to get the total systematic error.
  The errors for each particle species and setting are shown with results in Table~\ref{tbl:invslope_dndy}.

\section{FEED-DOWN FROM WEAK DECAYS}
  NA44 has no vertex detectors in the target region, so decay protons from some fraction of the short-lived particles such as strange baryons will be detected as protons in the spectrometer.
  This contamination of protons and antiprotons is not negligible.

  The same Monte-Carlo used for acceptance corrections was also used to estimate the weak decay yields from $\Lambda$, $\Sigma^+$ and $\Sigma^0$.
  Since the $\Sigma^0$ lifetime is short, its yield was included in the $\Lambda$ yield.
  The dN/dy for various particles near mid-rapidity in central collisions was taken from RQMD(v2.3).

\begin{figure}
    \includegraphics[width=8.5cm]{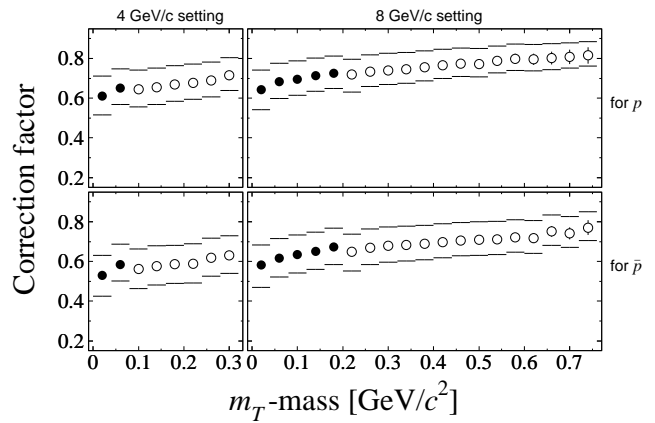}
    \caption{\label{fig:feeddown}
             The correction factors for proton and antiproton as a function of $m_T-$mass.
             The solid (open) circle corresponds to 44 (129) mrad setting.
             The horizontal bars above and below each point show the systematic errors from uncertainties in the hyperon/proton ratios. }
  \end{figure}

  The correction factor as a function of $m_T$ is
  \[
    C_{p}(m_{T}) = 
    \frac{N_{p_{\mbox{\tiny{orig}}}}(m_T)}
         {N_{p_{\mbox{\tiny{orig}}}}(m_T)+N_{p_{\Lambda}}(m_T)+N_{p_{\Sigma}}(m_T)}
  \]
  where $N_{p_{\mbox{\tiny{orig}}}}$ is the number of the original protons and $N_{p_{\Lambda}}$ and $N_{p_{\Sigma}}$ represent the protons from $\Lambda$ decays, and $\Sigma^+$ decays that track through the spectrometer.
  Figure~\ref{fig:feeddown} shows the correction factors as a function of $m_T-$mass.
  Since these depend on the ratio of $\Lambda$ and $\Sigma^+$ to proton, the production ratios were varied.
  The circles are the correction factor estimated from $\Lambda/p$ and $\Sigma/p$ ratios of RQMD, while the lines assume yield ratios larger or smaller by a factor 1.5.
  
  The correction procedure contributes to the systematic uncertainties on the inverse slopes of proton and antiproton by about 7\% for the 4 GeV/$c$ setting and about 3\% for the 8 GeV/$c$ setting.
  The uncertainties due to feed-down correction on dN/dy are approximately 15\% for protons and 19\% for antiprotons in both 4 GeV/$c$ and 8 GeV/$c$ settings.

\section{RESULTS}
  \label{section:results}
  \begin{figure*}
    \includegraphics[scale=0.66]{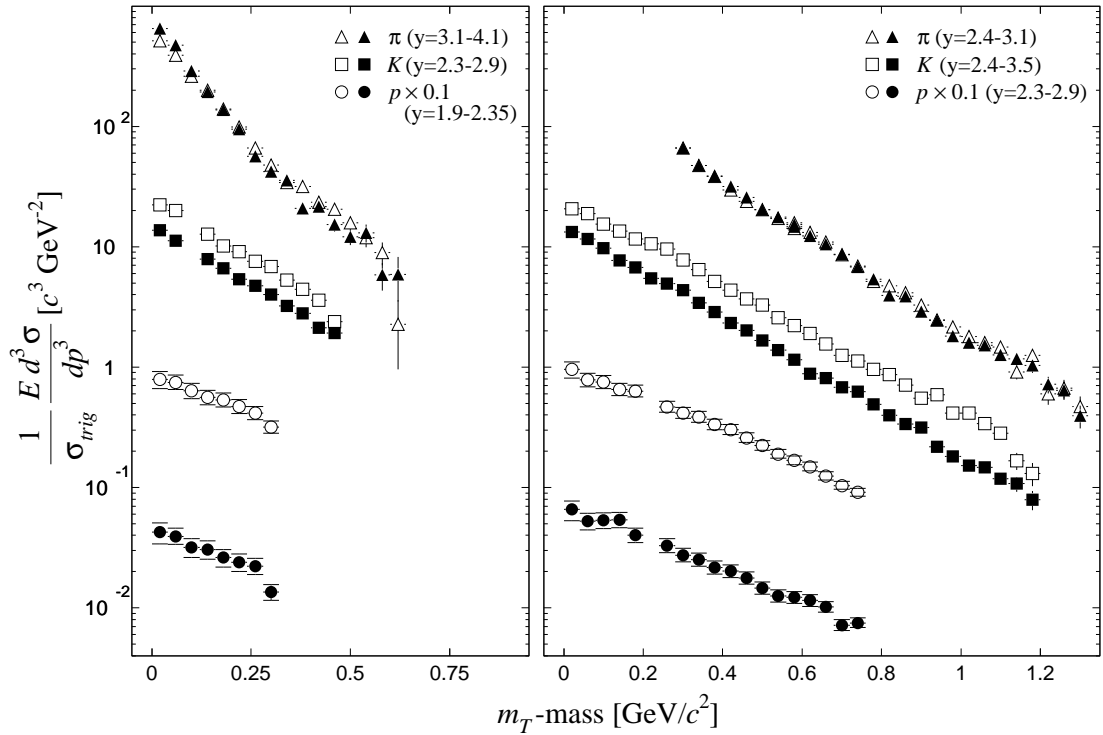}
    \caption{Invariant cross section of $\pi$, $K$, and $p$ in Pb+Pb collisions as a function of $m_T-$mass from the 4 GeV/$c$ and 8 GeV/$c$ setting.
             The open (solid) marker shows positive (negative) charged particle.
             The rapidity range for each particle is shown in the figure.
             The proton and antiproton distributions are corrected for feed-down.
             The horizontal bars above and below each point show systematic errors from the feed-down correction and correspond to the horizontal bars in Fig.~\protect\ref{fig:feeddown}.
             Note that the proton and antiproton are scaled by a factor of 0.1 in the vertical direction.
             }
    \label{fig:mtdist}
  \end{figure*}
  \begin{table*}
    \caption{\label{tbl:invslope_dndy}
             The inverse slope parameters and dN/dy with statistical errors (first) and systematic errors (second).
             The pion dN/dy is obtained from the data of 4 GeV/$c$ and 8 GeV/$c$ setting at both 44 mrad and 129 mrad.
             The values for proton and antiproton including the feed-down correction are shown as $p_c$ and $\bar{p}_c$, and systematic errors include errors from feed-down correction.
             }
    \begin{ruledtabular}
    \begin{tabular}{lccccc}
                &            & Fit range 
                                           & Inverse slope
                                                                   &  $\chi^{2}/DOF$       &        \\
                & rapidity   & $m_{T}-$mass  [GeV/$c^2$]
                                           & [MeV/$c^2$]
                                                                   & for inverse slope     & dN/dy  \\
      \hline
       $\pi^+$  & 2.4--3.1  & 0.28--1.20 & $207\pm3\pm_{2}^{3}$    & 92.3/25   &                      \\
                & 3.1--4.1  & 0.00--0.56 & $126\pm2\pm_{8}^{55}$   & 42.3/13   &                      \\
                & 2.4--4.1  &            &                         &           & $150\pm3\pm8$        \\
       $\pi^-$  & 2.4--3.1  & 0.28--1.20 & $201\pm3\pm_{1}^{2}$    & 63.0/25   &                      \\
                & 3.1--4.1  & 0.00--0.56 & $109\pm2\pm_{7}^{50}$   & 67.9/13   &                      \\
                & 2.4--4.1  &            &                         &           & $162\pm2\pm10$       \\
      $K^+$     & 2.3--2.9  & 0.00--0.32 & $221\pm9\pm_{2}^{10}$   & 17.1/9    & $24.0\pm0.2\pm2.1$   \\
                & 2.4--3.5  & 0.00--0.84 & $246\pm2\pm_{1}^{8}$    & 79.1/28   & $27.5\pm0.2\pm1.6$   \\
      $K^-$     & 2.3--2.9  & 0.00--0.32 & $224\pm7\pm1$           &  7.7/9    & $14.8\pm0.5\pm1.1$   \\
                & 2.4--3.5  & 0.00--0.84 & $228\pm2\pm_{1}^{2}$    & 49.2/28   & $15.4\pm0.5\pm1.0$   \\
      $p$       & 1.9--2.35 & 0.00--0.28 & $319\pm23\pm_{1}^{18}$  & 10.4/6    & $33.3\pm0.8\pm4.0$   \\
                & 2.3--2.9  & 0.00--0.68 & $296\pm5\pm_{7}^{17}$   & 74.1/16   & $34.7\pm1.0\pm2.2$   \\
      $\bar{p}$ & 1.9--2.35 & 0.00--0.28 & $303\pm35\pm_{7}^{34}$  &  9.0/6    & $1.74\pm0.01\pm0.33$ \\  
                & 2.3--2.9  & 0.00--0.68 & $300\pm9\pm_{1}^{7}$    & 18.5/16   & $2.67\pm0.05\pm0.26$ \\
      $p_c$     & 1.9--2.35 & 0.00--0.28 & $379\pm33\pm_{3}^{28}$  & 10.9/6    & $27.3\pm0.5\pm5.2$   \\
                & 2.3--2.9  & 0.00--0.68 & $327\pm6\pm_{10}^{28}$  & 94.5/16   & $28.5\pm0.2\pm4.6$   \\
    $\bar{p}_c$ & 1.9--2.35 & 0.00--0.28 & $360\pm53\pm_{15}^{47}$ &  8.7/6    & $1.43\pm0.04\pm0.39$ \\
                & 2.3--2.9  & 0.00--0.68 & $333\pm12\pm_{6}^{10}$  & 18.6/16   & $2.05\pm0.03\pm0.43$
    \end{tabular}
    \end{ruledtabular}
  \end{table*}
  The invariant cross sections measured by the NA44 spectrometer in Pb+Pb collisions are shown in Fig.~\ref{fig:mtdist}.
  They are plotted as a function of $m_T$ in the specified rapidity regions.
  The $m_T$ distribution is generally well described at these energies by
  \begin{equation}
    \label{eqn:mt_exponential}
    \frac{1}{\sigma_{trig}} \frac{E d^3 \sigma}{dp^3} = A e^{-(m_T-m) / T},
  \end{equation}
  where $A$ is a constant and $T$ is the inverse slope parameter.
  Table~\ref{tbl:invslope_dndy} lists the inverse slope parameters, the fit region, and the values of dN/dy.
  
  The slope parameters are consistent with previous NA44 results (Ref.~\cite{PRL_78_1997_2080}) within errors.
  The proton dN/dy is consistent with the values in Ref.~\cite{PL_B388_1996_431}.
  The statistics in our previous publication are $1/3$ to $3/4$ of the statistics used in this paper.
  We note that the NA44 collaboration published a paper on pion and kaon spectra focusing on ``strangeness enhancement''~\cite{PL_B471_1999_6}.
  This analysis used a different definition of the event fraction, based on the silicon multiplicity counter.
  The results are consistent within errors.
  
  \begin{figure}
    \includegraphics[width=8.7cm]{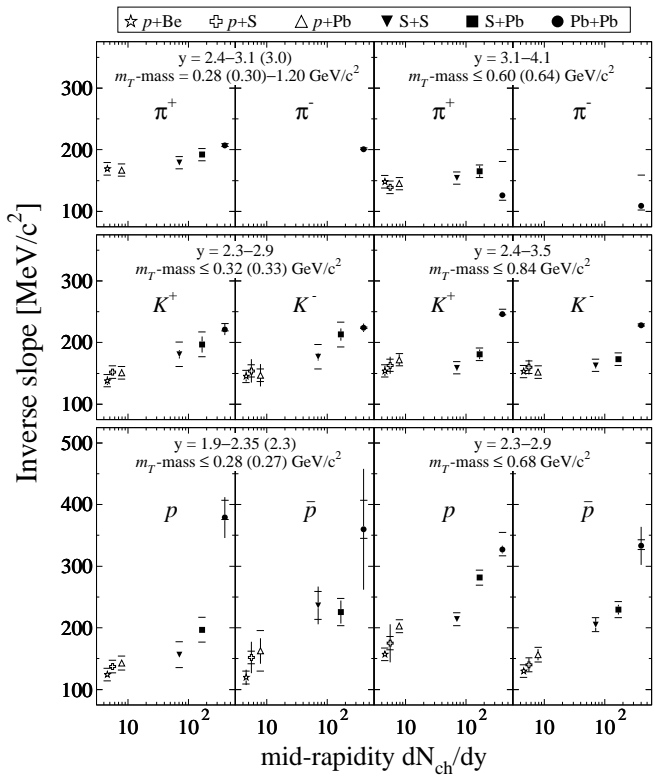}
    \caption{\label{fig:slope}
             Collision system dependence of charged pion, kaon, and proton inverse slope parameters.
             The slopes in $p$+A and S+A collisions are from Ref.~\protect\cite{PR_C57_1998_837,PR_C59_1999_328} .
             The $m_T$ and rapidity values in parenthesis are for $p$+A and S+A collisions.
             Statistical errors and systematic errors are shown as vertical and horizontal bars above and below each point, respectively.
            }
  \end{figure}

  The NA49 collaboration has reported dN/dy distributions of $\pi^{+}, \pi^{-}, K^{+}, K^{-}, p$, and $\bar{p}$ for the 5$\%$ most central Pb+Pb reactions~\cite{PRL_82_1999_2471,NP_A661_1999_45c,NP_A661_1999_362c,NP_A661_1999_383c}.
  These values are slightly higher than our data (top 3.7$\%$ event fraction).
  However, since they trigger on the energy of forward going spectators their selection of low impact parameter is better than ours.
  The top 5$\%$ NA49 data corresponds to mean impact parameter ${\langle}b{\rangle}$=2.2 fm~\cite{NP_A661_1999_362c}.
  On the other hand, the NA44 data corresponds to ${\langle}b{\rangle}$$\simeq$3 fm.
  Comparing similar impact parameter regions, the values of dN/dy are consistent with the NA49 results within the errors.
  
  We have obtained the invariant cross sections for pions, kaons, and protons in Pb+Pb collisions using data analysis procedures compatible with these used in Refs.~\cite{PR_C57_1998_837,PR_C59_1999_328}.
  Consequently, we can study the systematics of particle yields at CERN-SPS energy for many collision systems (450 GeV/$c$ $p$+A, 200 A GeV/$c$ S+A, and 158 A GeV/$c$ Pb+Pb).
  
  Figure~\ref{fig:slope} shows the inverse slope parameters for pions, kaons, and protons versus the $dN_{ch}/dy$ near central rapidity.
  All of the inverse slopes increase with $dN_{ch}/dy$ but this effect is more rapid for protons than for kaons and pions.
  The slopes of particles and antiparticles are equal except for protons have slightly larger inverse slopes than antiprotons for $p$+A and S+A collisions.
  While the uncertainties on the slope parameters for pions make a definitive statement difficult, the pion slope parameters in the region $m_T-$mass$<$0.6 GeV/$c^2$ in Pb+Pb are similar to, or less than, the slope in $p$+A and S+A collision systems.
  Perhaps, the lower values in the low $p_T$ region for Pb+Pb reflect a larger enhancement at low $p_T$.

  \begin{figure*}
    \includegraphics[width=17.5cm]{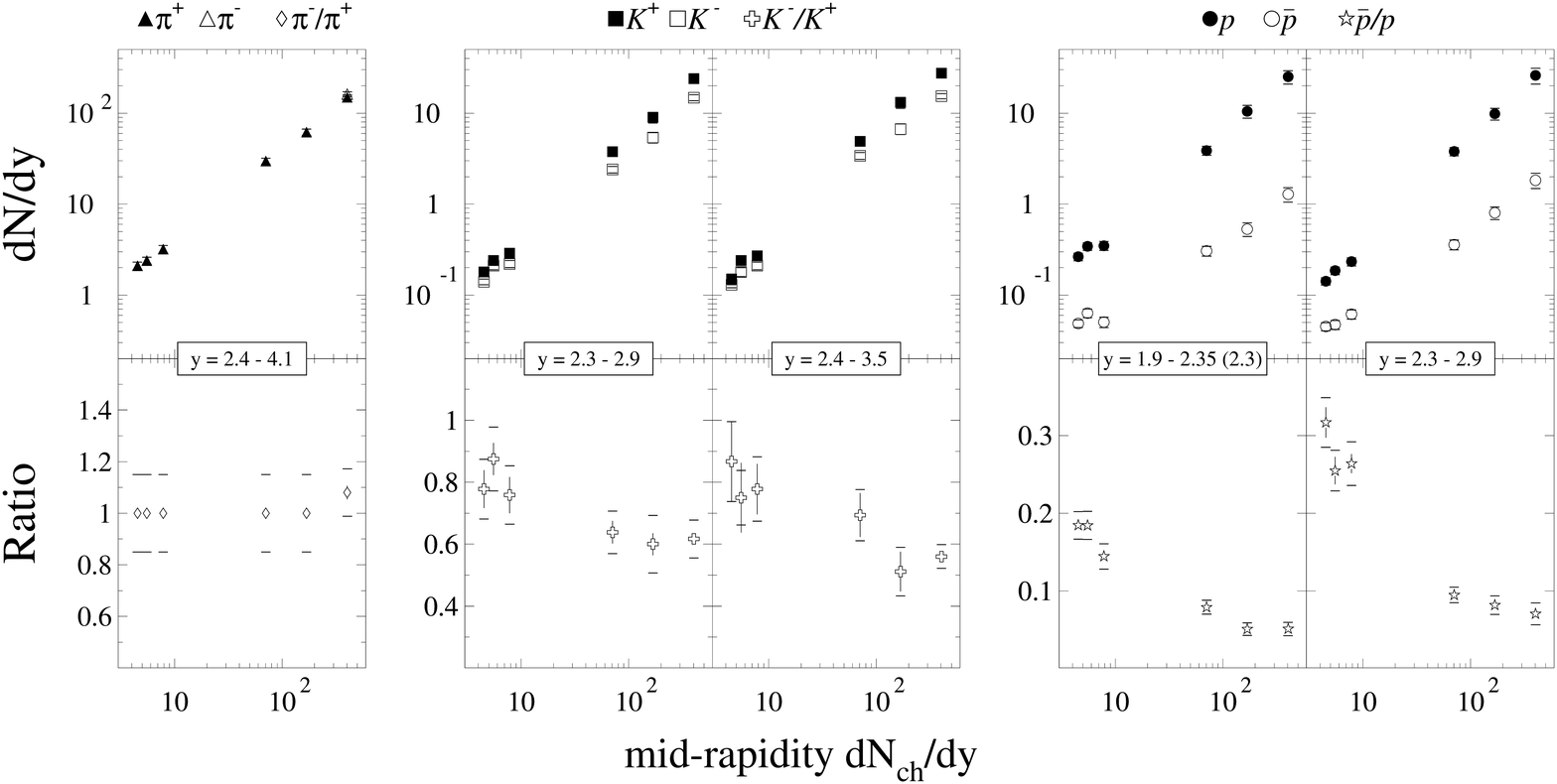}
    \caption{\label{fig:dndy}
             Collision system dependence of charged pion, charged kaon, proton, and antiproton dN/dy (top figures), and ratio of negative to positive particle (bottom figures) as a function of mid-rapidity charged particle rapidity density.
             The proton and antiproton dN/dy are corrected for feed-down.
             The values of dN/dy in $p$+A and S+A collisions are from Ref.~\protect\cite{PR_C57_1998_837,PR_C59_1999_328}.
             Statistical errors and systematic errors are shown as vertical and horizontal bars above and below each point, respectively.
            }
  \end{figure*}

  The kaon, proton and antiproton inverse slope parameters increase with collision system size slowly, but more rapidly than the pion inverse slope parameters.
  The proton inverse slopes are higher at mid-rapidity (y=2.3--2.9) than at more backward rapidities (y=1.9--2.3) in $p$+A and S+Pb collisions.
  
  Figure~\ref{fig:dndy} shows the collision system dependence of charged pion, charged kaon, proton, and antiproton yields, and ratios of negative to positive charged particle.
  The pion rapidity density in Pb+Pb collisions increases by a factor of $71\pm10$ compared to $p$+Be collisions, and by a factor $5.1\pm0.6$ compared to S+S collisions.
  dN/dy for $\pi^-$ and $\pi^+$ is compatible, within the errors for all collision systems.
  
  The mid-rapidity dN/dy of $K^+$ ($K^-$) in Pb+Pb collisions increases by a factor of $133\pm20$ ($105\pm10$) compared to $p$+Be collisions and $6.4\pm0.9$ ($6.2\pm0.8$) compared to S+S collisions.
  The system size dependence is stronger than for pions, i.e., the $K/\pi$ ratio increases from $p$+Be to Pb+Pb.
  
  Proton dN/dy increases with charged multiplicity at mid-rapidity more rapidly than antiprotons.
  In heavy-ion collisions, the $\bar{p}/p$ ratio decreases with system size, and falls more rapidly at mid-rapidity than at backward rapidity.
  This is exactly what is expected from increased stopping power, and therefore larger net baryon density, with larger colliding system size.

\section{DISCUSSION}

  In heavy-ion collisions, we use the notions of chemical freeze-out and thermal freeze-out to describe the moment when the inelastic and elastic interactions stop.
  In this section, we will describe the parameters extracted from the measured spectra and yields.
  
  \subsection{Transverse momentum distributions}
   
  The transverse momentum distributions of hadrons have been measured by several experimental groups at BNL-AGS and CERN-SPS including NA44~\cite{PRL_78_1997_2080}.
  The $m_T$ distributions are studied from the viewpoint of collective flow, with local thermal equilibrium at mid-rapidity, based on Refs.~\cite{PR_C48_1993_2462,PR_C54_1996_1390}.
  The NA44 collaboration has reported a thermal freeze-out temperature $T_{th}\approx 140$ MeV~\cite{PRL_78_1997_2080} in Pb+Pb collisions at $\sqrt{s_{NN}}\approx$ 18 GeV.
   Recently two-pion interferometry results have been combined with single particle spectra to measure the transverse flow and thermal freeze-out temperature~\cite{PRC_58_1998_2303,PRL_82_1999_2471}.
  These studies find $T_{th}\approx 120$ MeV~\cite{PRL_82_1999_2471} and 100 MeV~\cite{PRC_58_1998_2303}.
  Evaluating those reports equally, one concludes that the thermal freeze-out temperature is in the range 100--140 MeV.
  
  \subsection{Particle yields}
  
  If we assume a thermodynamic description is valid for these systems, then at chemical freeze-out, the density of different particles can be characterized by macroscopic parameters:
  the chemical freeze-out temperature and chemical potentials.
  Here, we adopt a chemical freeze-out model based on Ref.~\cite{PR_C59_1999_1637}, using the grand canonical ensemble.
  We assume no difference between $u$ and $d$ quark chemical potentials.
  Therefore, the hadron gas is described by a chemical freeze-out temperature ($T_{ch}$), light ($u$ and $d$) quark potential ($\mu_q$), strange quark potential ($\mu_s$), and strangeness saturation factor ($\gamma_s$).
  The density of a particle $i$ in the hadron gas is given by:
  \begin{equation}
    \label{eqn:rho}
    \rho_{i} = {\gamma_s}^{{\langle}s+\bar{s}{\rangle}_{i}} \frac{g_{i}}{2\pi^{2}} {T_{ch}}^3
               \left( \frac{m_i}{T_{ch}} \right)^2
               K_2(m_i/T_{ch}) \
               {\lambda_q}^{Q_i} \ {\lambda_s}^{s_i}
  \end{equation}
  where $m_i$ is the mass of the hadron $i$, $g_i$ is the number of spin-isospin degree-of-freedom, $K_2$ is the second-order modified Bessel function and,
  \[
    \lambda_q = \exp(\mu_q/T_{ch}),  \ \ \
    \lambda_s = \exp(\mu_s/T_{ch}).
  \]
  Note that we use the Boltzmann approximation for all hadrons except pions, where the Bose distribution is applied.
  The potential $\mu_q$ is for u/d/$\bar{\text{u}}$/$\bar{\text{d}}$ quarks, and $\mu_s$ is for s/$\bar{\text{s}}$ quarks.
  $Q_i$ and $s_i$ correspond to the net number of valence u/d quarks ($Q_{i}={\langle}u-\bar{u}+d-\bar{d}{\rangle}_{i}$), and s quark ($s_{i}={\langle}s-\bar{s}{\rangle}_{i}$) of particle species $i$, respectively.
  The factor $\gamma_s$ ($0{\le}\gamma_s{\le}1$) is introduced to allow of possibility of incomplete chemical equilibration for strange particles~\cite{PL_B262_1991_333}.
  $\gamma_s$=1 signifies full strangeness chemical equilibrium.
  The power factor of $\gamma_s$, ${\langle}s+\bar{s}{\rangle}_{i}$, is total number of s and $\bar{\mbox{s}}$ quarks in the particle.
  Ideally the system would live long enough that strange quarks would become fully equilibrated with the light quarks and $\gamma_s$ would be unity.
  However, calculations in Refs.~\cite{PR_C64_2001_024901,EPJ_C5_1998_143,JPG_25_1999_295} support $\gamma_s$$<$1.
  Therefore, we adopt $\gamma_s$ as a free parameter.
  The particle densities are computed by Eq.~(\ref{eqn:rho}) for the hadron gas including known resonances up to mass of 1.7 GeV/$c^2$, and the effect of broad resonance mass widths is taken into account~\cite{PR_C59_1999_1637}.
  The resonance decay to lower mass hadrons after chemical freeze-out is also taken into account.
  Therefore, the final particle densities for the pions, kaons, and protons are given by the sum of the direct production and the decays of the resonances.
  Here, we apply the above procedure for the particle yields observed near mid-rapidity by NA44 for $p$+A, S+A, and Pb+Pb collisions~\cite{PR_C57_1998_837,PR_C59_1999_328}.
  The model is fit to the data using MINUIT~\cite{MINUIT}.

  \begin{figure}[t]
    \includegraphics[width=8.5cm]{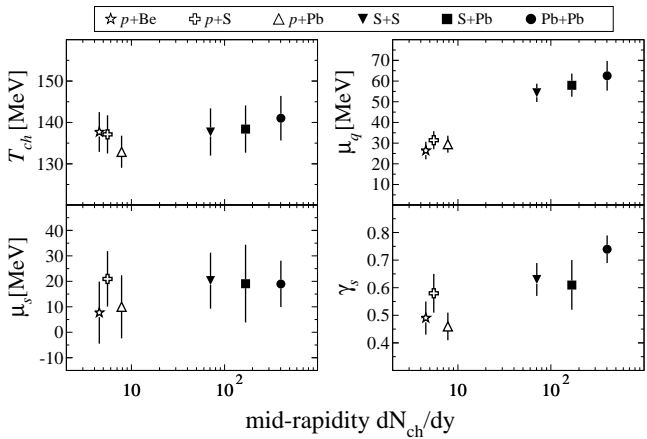}
    \caption{\label{fig:chemical}
             Collision system dependence of the chemical freeze-out parameters.
            }
  \end{figure}
  \begin{table}[t]
   \caption{\label{tbl:che_param}
            The macroscopic parameters for the description of the chemical freeze-out near central rapidity for Pb+Pb, S+A, and $p$+A collisions from the NA44 data.
            The values of dN/dy from Refs.~\protect\cite{PR_C57_1998_837,PR_C59_1999_328} are used to obtain the parameters in the $p$+A and S+A collisions.
           }
    \begin{ruledtabular}
    \begin{tabular}{lcccccc}
             &$T_{ch}$   & $\mu_q$  & $\mu_s$   &               &               \\
             &[MeV]      & [MeV]    & [MeV]     &$\gamma_s$     &$\chi^{2}/DOF$ \\
     \hline
      Pb+Pb  & 141$\pm$5 & 63$\pm$7 & 19$\pm$9  & 0.74$\pm$0.05 & 0.72/1             \\
      S+Pb   & 139$\pm$6 & 58$\pm$6 & 19$\pm$15 & 0.61$\pm$0.09 & 2.6$\times10^{-4}$/1\\
      S+S    & 138$\pm$4 & 54$\pm$4 & 20$\pm$11 & 0.63$\pm$0.06 & 3.8$\times10^{-5}$/1\\
      $p$+Pb & 133$\pm$4 & 30$\pm$4 & 10$\pm$12 & 0.46$\pm$0.05 & 1.0$\times10^{-5}$/1\\
      $p$+S  & 137$\pm$5 & 31$\pm$4 & 21$\pm$11 & 0.58$\pm$0.07 & 6.0$\times10^{-5}$/1\\
      $p$+Be & 138$\pm$5 & 26$\pm$4 &  8$\pm$12 & 0.49$\pm$0.06 & 3.9$\times10^{-4}$/1
    \end{tabular}
    \end{ruledtabular}
  \end{table}
  
  Figure~\ref{fig:chemical} and Table~\ref{tbl:che_param} shows the collision system dependence of the parameters.
  The chemical freeze-out temperatures ($T_{ch}$) are approximately 140 MeV in $p$+Be through Pb+Pb collisions: Within errors, there is no dependence on system size.
  Comparing to other results, the $T_{ch}$ is consistent with Ref.~\cite{JPG_25_1999_295} ($T_{ch}$=133--144 MeV) and smaller than Ref.~\cite{PR_C64_2001_024901} ($T_{ch}$=158$\pm$3 MeV) and Ref.~\cite{PL_B465_1999_15} ($T_{ch}$=168$\pm$2.4 MeV).
  As reported in Refs.~\cite{PR_C59_1999_1637,JPG27_2001_589} there is a rapidity dependence of the chemical freeze-out parameters.
  Those calculations show that a lower value of the temperature is obtained when only mid-rapidity data are used.
  The limited number of particle species used in this analysis may also influence the value of $T_{ch}$.

  The values of $\mu_q$ increase with the size of the collision system.
  Since the chemical freeze-out temperature is independent of collision systems ($p$+A and A+A), the system in $p$+A collisions has smaller $\mu_q/T_{ch}$, that is, smaller baryon density than A+A collisions.
  $\mu_s$ is 10$-$20 MeV for $p$+A through Pb+Pb.
  $\mu_s$ should be zero in a equilibrated quark gluon plasma and on the phase boundary of a hadron gas~\cite{PL_B262_1991_333,PR_C37_1988_1452,PR_D51_1995_1086}.
  Consequently, the small value we obtain may suggest that chemical freeze-out takes place near the phase transition.
  
  Finally we find that $\gamma_s$ increases logarithmically with $dN_ch/dy$ from $0.49\pm0.06$ in $p$+Be collisions to $0.74\pm0.05$ in Pb+Pb.
  Thus, we seem to be approaching full chemical equilibrium for strange particles in heavy-ion collisions.
  The use of the grand canonical ensemble for analysis of strange particle production in $p$+A collisions does not take into account phase space effects arising from production of $s$ and $\bar{s}$ pairs.
  Consequently, $\gamma_s$ is expected to be lower than in heavy-ion collisions where more strange quark pairs are produced~\cite{JPG27_2001_413,NP_A698_2002_94}.

\section{CONCLUSION}
  The NA44 experiment has measured single particle distributions for charged pions, kaons, and protons as functions of transverse mass ($m_T$) near mid-rapidity in 158 A GeV/$c$ Pb+Pb collisions.
  We studied particle ratios near mid-rapidity from the viewpoint of chemical freeze-out for $p$+A (A = Be, S, and Pb), S+A (A = S, Pb), and Pb+Pb collisions.
  We find that chemical freeze-out occurs at a temperature of $\approx$140 MeV for all collisions.
  The values of the light quark chemical potential ($\mu_q$) are approximately 30 MeV in $p$+A and 54 to 63 MeV in S+S to Pb+Pb collisions.
  The trend of $\mu_q$ shows that the baryon stopping power increases with collision system size.
  The values of the strange quark chemical potential are about 10 to 20 MeV.
  The strangeness saturation factor is 0.61 to 0.75 in heavy-ion collisions and 0.46 to 0.58 in $p$+A,
  indicating that $p$+A collisions are further away from strangeness equilibration than A+A collisions.

\begin{acknowledgments}
  The NA44 collaboration wishes to thank the staff of the CERN PS-SPS accelerator complex for their excellent work. 
  We are also grateful for support given by 
  the Danish Natural Science Research Council;
  the Japanese Society for the Promotion of Science; 
  the Ministry of Education, Science and Culture, Japan; 
  the Swedish Science Research Council;
  the Austrian Fond f\"{u}r F\"{o}rderung der Wissenschaftlichen Forschung; 
  the National Science Foundation, and the US Department of Energy.

\end{acknowledgments}

\bibliography{pbpbsingle}

\end{document}